# An Earth-sized planet with an Earth-like density

Francesco Pepe[1], Andrew Collier Cameron[2], David W. Latham[3], Emilio Molinari[4,5], Stéphane Udry[1], Aldo S. Bonomo[6], Lars A. Buchhave[3,7], David Charbonneau[3], Rosario Cosentino[4,8], Courtney D. Dressing[3], Xavier Dumusque[3], Pedro Figueira[9], Aldo F. M. Fiorenzano[4], Sara Gettel[3], Avet Harutyunyan[4], Raphaëlle D. Haywood[2], Keith Horne[2], Mercedes Lopez-Morales[3], Christophe Lovis[1], Luca Malavolta[10,11], Michel Mayor[1], Giusi Micela[12], Fatemeh Motalebi[1], Valerio Nascimbeni[11], David Phillips[3], Giampaolo Piotto[10,11], Don Pollacco[13], Didier Queloz[1,14], Ken Rice[15], Dimitar Sasselov[3], Damien Ségransan[1], Alessandro Sozzetti[6], Andrew Szentgyorgyi[3] & Christopher A. Watson[16]

[1]Observatoire Astronomique de l'Université de Genève, 51 chemin des Maillettes, 1290 Versoix, Switzerland.

[2]SUPA, School of Physics and Astronomy, University of St Andrews, North Haugh, St Andrews, Fife KY16 9SS, UK.

[3]Harvard-Smithsonian Center for Astrophysics, 60 Garden Street, Cambridge, Massachusetts 02138, USA.

[4]INAF - Fundación Galileo Galilei, Rambla José Ana Fernandez Pérez 7, 38712 Breña Baja, Spain.

[5]INAF - IASF Milano, via Bassini 15, 20133 Milano, Italy.

[6]INAF - Osservatorio Astrofisico di Torino, via Osservatorio 20, 10025 Pino Torinese, Italy.

[7]Centre for Star and Planet Formation, Natural History Museum of Denmark, University of Copenhagen, DK-1350 Copenhagen, Denmark.

[8]INAF - Osservatorio Astrofisico di Catania, via S. Sofia 78, 95125 Catania, Italy.

[9]Centro de Astrofísica, Universidade do Porto, Rua das Estrelas, 4150-762 Porto, Portugal.

[10]Dipartimento di Fisica e Astronomia "Galileo Galilei", Universita' di Padova, Vicolo dell'Osservatorio 3, 35122 Padova, Italy.

[11]INAF - Osservatorio Astronomico di Padova, Vicolo dell'Osservatorio 5, 35122 Padova, Italy.

[12]INAF - Osservatorio Astronomico di Palermo, Piazza del Parlamento 1, 90124 Palermo, Italy.

[13]Department of Physics, University of Warwick, Gibbet Hill Road, Coventry CV4 7AL, UK.

[14]Cavendish Laboratory, J. J. Thomson Avenue, Cambridge CB3 0HE, UK.

[15]SUPA, Institute for Astronomy, Royal Observatory, University of Edinburgh, Blackford Hill, Edinburgh EH93HJ, UK.

[16]Astrophysics Research Centre, School of Mathematics and Physics, Queen's University Belfast, Belfast BT7 1NN, UK.

**Recent analyses[1–4] of data from the NASA Kepler spacecraft[5] have established that planets with radii within 25 per cent of Earth's ($R_\oplus$) are commonplace throughout the Galaxy, orbiting at least 16.5 per cent of Sun-like stars[1]. Because these studies were sensitive to the sizes of the planets but not their masses, the question remains whether these Earth-sized planets are indeed similar to the Earth in bulk composition. The smallest planets for which masses have been accurately determined[6,7] are Kepler-10b ($1.42R_\oplus$) and Kepler-36b ($1.49R_\oplus$), which are both significantly larger than the Earth. Recently, the planet Kepler-78b was discovered[8] and found to have a radius of only $1.16R_\oplus$. Here we report that the mass of this planet is 1.86 Earth masses. The resulting mean density of the planet is 5.57 g cm$^{-3}$, which is similar to that of the Earth and implies a composition of iron and rock.**

Every 8.5 h, the star Kepler-78 (first known as TYC 3147-188-1 and later designated KIC 8435766) presents to Earth a shallow eclipse consistent[8] with the passage of an orbiting planet with a radius of $1.16 \pm 0.19 R_\oplus$. A previous study[8] demonstrated that it was very unlikely that these eclipses were the result of a massive companion either to Kepler-78 itself or to a fainter star near its position on the sky. Judging from the absence of ellipsoidal light variations[8] of the star, the upper limit on the mass of the planet is 8 Earth masses ($M_\oplus$). In addition to its diminutive size, the planet Kepler-78b is interesting because the light curve recorded by the Kepler spacecraft reveals the secondary eclipse of the planet behind the star as well as the variations in the light received from the planet as it orbits the star and presents different hemispheres to the observer. These data enabled constraints[8] to be put on the albedo and temperature of the planet. A direct measurement of the mass of Kepler-78b would permit an evaluation of its mean density and, by inference, its composition, and motivated the study we describe here.

The newly commissioned HARPS-N[9] spectrograph is the Northern Hemisphere copy of the HARPS[10] instrument, and, like HARPS, HARPS-N allows scientific observations to be made alongside thorium–argon emission spectra for wavelength calibration[11]. HARPS-N is installed at the 3.57-m Telescopio Nazionale Galileo at the Roque de los Muchachos Observatory, La Palma Island, Spain. The high optical efficiency of the instrument enables a radial-velocity precision of 1.2 m s$^{-1}$ to be achieved in a 1-h exposure on a slowly rotating late-G-type or K-type dwarf star with $m_v = 12$. By observing standard stars of known radial velocity during the first year of operation of HARPS-N, we estimated it to have a precision of at least 1 m s$^{-1}$, a value which is roughly half the semi-amplitude of the signal expected for Kepler-78b should the planet have a rocky composition. We began an intensive observing campaign (Methods) of Kepler-78 ($m_v = 11.72$) in May 2013, acquiring HARPS-N spectra of 30-min exposure time and an average signal-to-noise ratio of 45 per extracted pixel at 550 nm (wavelength bin of 0.00145 nm). From these high-quality spectra, we estimated[12,13] the stellar parameters of Kepler-78 (Methods and Extended Data Table 1). Our estimate of the stellar radius, $R_* = 0.737^{+0.034}_{-0.042} R_e$, is more accurate than previously known[8] and allows us to refine the estimate of the planetary radius.

In Supplementary Data, we provide a table of the radial velocities, the Julian dates, the measurement errors, the line bisector of the cross-correlation function, and the Ca II H-line and K-line activity indicator[14], $\log(R'_{HK})$. The radial velocities (Fig. 1) show a scatter of 4.08 m s$^{-1}$ and a peak-to-trough variation of 22 m s$^{-1}$, which exceeds the estimated average internal (photon-noise) precision, of 2.3 m s$^{-1}$. The excess scatter is probably due to star-induced effects including spots and changes in the convective blueshift associated with variations in the stellar activity. These effects may cause an apparent signal at the stellar rotational period and its first and second harmonics. To separate this signal from that caused by the planet, we proceeded to estimate the rotation period of the star from the de-trended light curve from Kepler (Methods). Our estimate, of ~12.6 d, is consistent with a previous estimate[8]. The power spectral density of the de-trended light curve also shows strong harmonics at respective periods of 6.3 and 4.2 d. We note that these timescales are much longer than the orbital period of the planet.

To estimate the radial-velocity semi-amplitude, $K_p$, due to the planet, we proceeded under the assumption that $K_p$ was much larger than the change in radial velocity arising from stellar activity during a single night. This is a reasonable assumption, because a typical 6-h observing sequence spans only 2% of the stellar rotation cycle. Furthermore, the relative phase between the stellar signal and the planetary signal changes continuously, and so we expect that the contribution from stellar activity should average out over the three-month observing period. Assuming a circular orbit, we modelled the radial velocity of the $i$th observation (gathered at time $t_i$ on night $d$) as $v_{i,d} = v_{0,d} - K_p \sin(2\pi(t_i - t_0)/P)$, where $t_0$ is the epoch of mid-transit and $P$ is the orbital period, each held fixed at the values derived from the photometry, and $v_{0,d}$ is the night-$d$ zero point. We solved for the values of $K_p$ and $v_{0,d}$ by minimizing the $\chi^2$ function, assuming white noise and weighting the data according to the inverse variances derived from the HARPS-N noise model. This is a technique similar to the one used in a recent study[15] of another low-mass transiting planet, CoRoT-7b.

With this technique, we measure a preliminary value of $K_p = 2.08 \pm 0.32$ m s$^{-1}$, implying a detection significance of 6.5$\sigma$. We confirmed that the radial-velocity signal is consistent with the photometric transit ephemeris by repeating the $\chi^2$ optimization over a grid of orbital frequencies and times of mid-transit (Fig. 2a). This confirms that the values of $P$ and $t_0$ from the Kepler light curves coincide with the lowest $\chi^2$ minimum in the resulting period–phase diagram.

The Kepler light curve evolves on a timescale of weeks. We therefore expect the stellar-activity-induced radial-velocity signals to remain coherent on the same timescale. To explore this, we fitted the radial velocities with a sum of Keplerian models at different periods, one of which we expect to correspond to the planetary orbit and the others (which we left as free parameters) to represent the effects of stellar activity. Using the Bayesian information criterion as our discriminant (Methods), we found that a model with three Keplerians was sufficient to explain the data. The 4.2-d period of the second Keplerian corresponds to the second harmonic of the photometrically determined stellar rotation period. The period of 10 d for the third Keplerian also appears as a prominent peak in the generalized Lomb–Scargle analysis[16] of the radial-velocity data. Strong peaks at similar periods are also present in periodograms of the Ca II HK activity indicator, the line bisector and the full-width at half-maximum of the cross-correlation function[11] (Extended Data Fig. 1). We conclude that both the 4.2-d and 10-d signals probably have stellar causes.

The best three-Keplerian fit of the data yields an estimate of $K_p = 1.96 \pm 0.32$ m s$^{-1}$ and residuals with a dispersion of 2.34 m s$^{-1}$, very close to the internal noise estimates. In Fig. 2b, we show the phase-folded radial velocities after removal of the stellar components, plotted along with the best-fit Keplerian at the planetary orbital period. The orbital parameters are given in Table 1. Having settled on the three-component model, we estimated the planetary mass and density by conducting a Markov chain Monte Carlo (MCMC) analysis (Methods). In this analysis, we adopted previously published[8] values of $P$ and $t_0$ as Gaussian priors. We replaced the planet radius, $R_p$, with the published estimate[8] of the ratio $k = R_p/R_*$ and our determination of $R_*$. The planetary radius then becomes an output of the MCMC analysis. By adopting the mode of the distributions, we find a planet mass of $m_p = 1.86^{+0.38}_{-0.25} M_\oplus$, a radius of $R_p = 1.173^{+0.159}_{-0.089} R_\oplus$ and a planet density of $\rho_p = 5.57^{+3.02}_{-1.31}$ g cm$^{-3}$. These estimates include the contribution from the uncertainty in the stellar mass. The uncertainty in the density is dominated by the uncertainty in $k$. Our values for $m_p$ and $\rho_p$ are consistent with those from an independent study[17].

In terms of mass, radius and mean density, Kepler-78b is the most similar to the Earth among the exoplanets for which these quantities have been determined. We plot the mass–radius diagram in Fig. 3. By comparing our estimates of Kepler-78b with theoretical models[18] of internal composition, we find that the planet has a rocky interior and most probably a relatively large iron core (perhaps comprising 40% of the planet by mass). We note that in the part of the mass–radius diagram where Kepler-78b lies, there is a general agreement between models and little or no degeneracy. The extreme proximity of the planet to its star, resulting in a high surface temperature and ultraviolet irradiation, would preclude there being a low-molecular-weight atmosphere: any water or volatile envelope that Kepler-78b might have had at formation should have rapidly evaporated[19]. Kepler-78b is also similar to larger high-density, hot exoplanets (Kepler-10b (ref. 6), Kepler-36b (ref. 7) and CoRoT-7b (ref. 20)), in that in the mass–radius diagram it is not below the lower envelope of mantle-stripping models[21] that tend to enhance the fraction of the planet's iron core. At present, Kepler-78b is the extrasolar planet whose mass, radius and likely composition are most similar to those of Earth. However, it differs from Earth notably in its very short orbital period and correspondingly high temperature.

The observations of Kepler-78 have shown the potential of the much-anticipated HARPS-N spectrograph. It will have a crucial role in the characterization of the many Kepler planet candidates with radii similar to that of Earth. By acquiring and analysing a large number of precise radial-velocity measurements, we can learn whether Earth-sized planets (typically) have Earth-like densities (and, by inference, Earth-like compositions), or whether even small planets have a wide range of compositions, as has recently been established[22,23] for their larger kin.

**Acknowledgements**

This Letter was submitted simultaneously with the paper by Howard *et al.*[17]. Both papers are the result of a coordinated effort to carry out independent radial-velocity observations and studies of Kepler-78. Our team greatly appreciates the spirit of this collaboration, and we sincerely thank A. Howard and his team for the collegial work. We wish to thank the technical personnel of the Geneva Observatory, the Astronomical Technology Centre, the Smithsonian Astrophysical Observatory and the Telescopio Nazionale Galileo for their enthusiasm and competence, which made the HARPS-N project possible. The HARPS-N project was funded by the Prodex Program of the Swiss Space Office, the Harvard University Origins of Life Initiative, the Scottish Universities Physics Alliance, the University of Geneva, the Smithsonian Astrophysical Observatory, the Italian National Astrophysical Institute, the University of St Andrews, Queen's University Belfast and the University of Edinburgh. P.F acknowledges support from the European Research Council/European Community through the European Union Seventh Framework Programme, Starting Grant agreement number 239953, and from the Fundação para a Ciência e a Tecnologia through grants PTDC/CTE-AST/098528/2008 and PTDC/CTE-AST/098604/2008. The research leading to these results received funding from the European Union Seventh Framework Programme (FP7/2007-2013) under grant agreement number 313014 (ETAEARTH).

**Author Contributions** The underlying observation programme was conceived and organized by F.P., A.C.C., D.L., C.L, D.Se, S.U.. and E.M.. Observations with HARPS-N@TNG were carried out by A.C.C., A.S.B., D.C., R.C., C.D.D., X.D., P.F, A.F.M.F, S.G., A.H., R.D.H., M.L.-M., V.N., D. P., D.Q., K.R., A.So., A.Sz.. and C.A.W. The data-reduction pipeline was adapted and updated by C.L., who also implemented the correction for Charge-Transfer Efficiency-errors and the automatic computation of the activity indicator log($R'_{HK}$). M.L.-M. and S.G. independently computed the $S$-index and log($R'_{HK}$) values. A.C.C., D.Se., A.S.B. and X.D. analysed the data using the offset-correction method. A.C.C. and D.S. re-analysed the data for the determination of the stellar rotational period based on the Kepler light curve. P.F. investigated for possible correlations between the radial velocities and the line bisector. L.A.B. conducted the stellar parameter classification analysis for the re-determination of the stellar parameters based on HARPS-N spectra. An independent determination of the stellar parameters was conducted by L.M. by analysis of the cross-correlation function. D.Se. compared many different models to fit the observed data and selected the most appropriate by using the Bayesian information criterion. D.Se. performed a detailed MCMC analysis for the determination of the planetary parameters, with contributions also from A.S.B. and A.So.. F.P. was the primary author of the manuscript, with important contributions by D.C., D.Se., D.Sa, C.A.W., K.R. and C.L.. All authors are members of the HARPS-N Science Team and have contributed to the interpretation of the data and the results.

**Author Information** Correspondence and requests for materials should be addressed to F.P. ([francesco.pepe@unige.ch](francesco.pepe@unige.ch)).

## METHODS

**Photometric determination of stellar rotational period of Kepler-78**

In ref. 8, the Kepler light curve of Kepler-78 was analysed and was de-trended using the PDC-MAP algorithm (Extended Data Fig. 2), which preserves stellar variability[24,25]. The light curve displays clear rotational modulation with a peak-to-valley amplitude that varies between 0.5% and 1.5%, and a period of 12.6 ± 0.3 d. We confirmed the rotational period by computing the autocorrelation function of the PDC-MAP light curve: Using a fast Fourier transform we compute the power spectral density from which the autocorrelation function (ACF) is obtained using the inverse transform. We immediately derive a rotational period of 12.6 d (Extended Data Fig. 3a). The amplitudes of successive peaks decay on an e-folding timescale of about 50 d, which we attribute to the finite lifetimes of individual active regions. The power density distribution in Extended Data Fig. 3b finally shows a peak at the stellar rotational period as well as at its first and second harmonics. The main signal at period $P = 0.355$ d of Kepler-78b, as well as its harmonics, are easily identified at shorter periods.

**HARPS-N observations and stellar parameters**

To explore the feasibility of the programme, we performed five hours of continuous observations during a first test night in May 2013. This test night allowed us to determine the optimum strategy and to verify whether the measurement precision was consistent with expectations. Indeed, 12 exposures, each of 30 min, led to an observed dispersion of the order of 2.5 m s$^{-1}$, close to the expected photon noise. We therefore decided to dedicate six full HARPS-N nights to the observation of Kepler-78b in June 2013. Given the excellent stability of the instrument (typically less than 1 m s$^{-1}$ during the night) and the faintness of the star, we observed without the simultaneous reference source[10,11] that usually serves to track potential instrumental drifts. Instead, the second fibre of the spectrograph was placed on the sky to record possible background contamination during cloudy moonlit nights. Owing to excellent astroclimatic conditions, we gathered a total of 81 exposures, each of 30 min, free of moonlight contamination and with an average signal-to-noise ratio (SNR) of 45 per extracted pixel at a wavelength of $\lambda = 550$ nm. An extracted pixel covers a wavelength bin of 0.000145 nm.

A first analysis of these observations confirmed the presence of the planetary signal. However, it also confirmed that, as suggested in ref. 8, the stellar variability induces radial-velocity variations much larger than the planetary signal, although on very different timescales. To consolidate our results and improve the precision of our planetary-mass measurement, we decided to perform additional observations during the months of July and August 2013. We preferred, however, to observe Kepler-78 only twice per night, around quadrature (at maximum and minimum expected radial velocity), to minimize observing time and to maximize the information on the amplitude. This strategy allowed us to determine the (low-frequency) stellar contribution as the sum of the two nightly measurements and the (high-frequency) planetary signal as the difference between them. We finally obtained a total of 109 high-quality observations over three months, with an average photon-noise-limited precision of 2.3 m s$^{-1}$.

The large number of high-SNR spectra gathered by HARPS-N allowed us to re-determine the stellar parameters using the stellar parameter classification (SPC) pipeline[12]. Each high-resolution spectrum ($R = 115{,}000$) yields an average SNR per resolution element of 91 in the MgB region. The weighted average of the individual spectroscopic analyses resulted in final stellar parameters of $T_{\mathrm{eff}} = 5058 \pm 50$ K, $\log(g) = 4.55 \pm 0.1$, $[\mathrm{m/H}] = -0.18 \pm 0.08$ and $v\sin(i) = 2 \pm 1$ km s$^{-1}$, in agreement, within the uncertainties, with the discovery paper. The value for $v\sin(i)$ is, however, poorly determined by SPC. Therefore, we adopted an internal calibration based on the full-width at half-maximum of the cross-correlation functions to compute the projected rotational velocity, which yielded $v\sin(i) = 2.8 \pm 0.5$ km s$^{-1}$. We note that, assuming spin–orbit alignment, the rotational velocity and our estimate of the stellar radius yield a rotational period of 13 d. This value is in agreement with the stellar rotational period determined from photometry.

The stellar parameters from SPC[12] have been input to the Yonsei–Yale stellar evolutionary models[13] to estimate the mass and radius of the host star. We obtain $M_* = (0.758 \pm 0.046)\,M_\text{e}$ for the stellar mass and $R_* = 0.737^{+0.034}_{-0.042}\,R_\text{e}$ for the radius, in agreement, within the uncertainties, with the discovery paper. The Ca II HK activity indicator is computed by the online and automatic data-reduction pipeline, which gives an average value of $\log(R'_\text{HK}) = -4.52$ when using a colour index $B-V = 0.91$ for Kepler-78. The stellar parameters are summarized in Extended Data Table 1.

**Radial-velocity model selection**

It is interesting to note that the signature of Kepler-78b can be retrieved despite the large stellar signals superimposed on the radial velocity induced by the planet. To demonstrate this, we adjusted the data with a simple model consisting of a cosine and the star's systemic velocity, while fixing the period and time of transit to the published values[8]. We compared the results of this simple model with a simple constant using the Bayesian information criterion[26–28] (BIC). We derived the relative likelihood of the two models, also called the evidence ratio, to be $\text{e}^{-(1/2)\Delta\text{BIC}_i} = 4\times10^{-10}$. This first estimate tells us that our cosine model is much superior to the simple constant. In other words, we can say that we have a clear detection of a signal of semi-amplitude $K_\text{p} = 1.88 \pm 0.47$ m s$^{-1}$. Although certainly biased owing to the lack of stellar activity de-trending, the result provides a confirmation of the existence of Kepler-78b and a first estimation of its mass.

To model the stellar signature, we followed two different approaches. The first one consists of removing any stellar effect occurring on a timescale longer than 2 d by adjusting nightly offsets to the data. This method has the main advantage of not relying on any analytical model and it overcomes the difficulty of modelling non-stationary processes that often characterize stellar activity. The approach is also well suited to our problem because the period of the planet is very short. Its only drawback comes from the large number of additional parameters (21 offsets, one per night), which is a direct consequence of our observing strategy. The second approach consists of modelling the stellar activity as a set of sinusoids or Keplerians. This approach makes sense provided that spot groups and plages are coherent on a timescale similar to the radial-velocity observation time span. For Kepler-78, the ACF of the light curve shows a 1/e de-correlation of ~50 d (Extended Data Fig. 3a), which compares well with the 97-d time span of the HARPS-N observations.

In total, we studied a series of more than 30 different models of different complexity. We have compared these models using the BIC[28] evidence ratio, ER, and the BIC weight, $w$, to find the best few models:

$$w_i = \frac{e^{-(1/2)\Delta BIC_i}}{\sum_i e^{-(1/2)\Delta BIC_i}}$$

Of all the models we considered, two are statistically much more significant. They consist of modelling the stellar activity as a sum of periodic signals. The best model, with a BIC weight of 0.71, foresees three Keplerians including the planetary signal ($P_1$ = 0.355 d, $P_2$ = 4.2 d and $P_3$ = 10.04 d; $e_1$, $e_2$ = 0). The second best contains four Keplerians ($P_1$ = 0.355 d, $P_2$ = 4.2 d, $P_3$ = 6.5 d and $P_4$ = 23–58 d; all eccentricities $e_i$ = 0). De-trending of the stellar activity using nightly offsets shows much weaker evidence ratios. We therefore retained model 5 (Extended Data Table 2), which consists of one Keplerian describing the planet Kepler-78b and two additional 'signals', a sinusoid of period $P_2$ = 4.2 d and a slightly eccentric Keplerian with $P_3$ = 10.04 d. In Extended Data Table 3, we present the distribution of the parameters of our best model as resulting from the MCMC analysis (see next section). Furthermore, Extended Data Fig. 4 shows the periodogram of the radial-velocity residuals after subtracting the stellar components. The planetary signal is now detected with a false-alarm probability significantly lower than 1%.

**MCMC analysis**

To retrieve the marginal distribution of the true mass of the planet and its density, we carry out an MCMC analysis based on the model selection process described in the previous section. We sample the posterior distributions using an MCMC with the Metropolis-Hastings algorithm. Because the model is very well constrained by the data, the MCMC starts from the solution corresponding to the maximum likelihood, and the MCMC parameter steps correspond to the standard deviation of the adjusted parameters. An acceptance rate of 25% is chosen. To obtain the best possible end result, we take as priors the transit parameters of Kepler-78b (ref. 8). Symmetric distributions are considered to be Gaussian, whereas asymmetric ones, such as that of the orbital inclination, are modelled by split-normal distributions using the published value of the mode of the distribution. We re-derive the radius of the planet using our improved stellar radius estimation and the planet-to-star radius ratio from the Kepler photometry[8]. All other parameters have uniform priors except for the period $P$, for which a modified Jeffrey's prior is preferred[29]. We use $\sqrt{e}\cos(\omega)$ and $\sqrt{e}\sin(\omega)$ as free parameters, which translate into a uniform prior in eccentricity[30]. The mean longitude, $\lambda_0$, computed at the mean date of the observing campaign, is also preferred as a free parameter. It has the advantage of not being degenerate for low eccentricities, whereas our choice for the reference epoch, $T_0$, reduces correlations between adjusted parameters. In this analysis, the MCMC has 2,000,000 iterations and converges after a few hundred iterations. The ACF of each parameter is computed to estimate the typical correlation length of our chains and to estimate a sampling interval to build the final statistical sample. All ACFs have a very short decay (1/e decay after 100 iterations and 1/100 decay after 300 iterations) and present no correlations on a larger iteration lag. We build our final sample using the 1/e-decay iteration lag, which is a good compromise between the size of the statistical sample and its de-correlation value. The final statistical samples consist of 20,000 elements, from which orbital elements and confidence intervals are derived. The resulting orbital elements for Kepler-78b are listed in Extended Data Table 3. The results for the mass, radius and density of the planet are given in Extended Data Table 4, and the distributions for the mass and the density are plotted in Extended Data Fig. 5. These distributions are smoothed for better rendering.

**REFERENCES**

1. Fressin, F. et al. The false positive rate of Kepler and the occurrence of planets. *Astrophys. J.* **766,** 81–100 (2013)
2. Petigura, E. A., Marcy, G. W. & Howard, A. W. A plateau in the planet population below twice the size of Earth. *Astrophys. J.* **770,** 69–89 (2013)
3. Swift, J. J. et al. Characterizing the cool KOIs. IV. Kepler-32 as a prototype for the formation of compact planetary systems throughout the galaxy. *Astrophys. J.* **764,** 105–118 (2013)
4. Dressing, C. D. & Charbonneau, D. The occurrence rate of small planets around small stars. *Astrophys. J.* **767,** 95–114 (2013)
5. Batalha, N. M. et al. Planetary candidates observed by Kepler. III. Analysis of the first 16 months of data. *Astrophys. J.* **204** (suppl.)**,** 24–44 (2011)
6. Batalha, N. M. et al. Kepler's first rocky planet: Kepler-10b. *Astrophys. J.* **729,** 27–47 (2011)
7. Carter, J. A. et al. Kepler-36: a pair of planets with neighboring orbits and dissimilar densities. *Science* **337,** 556–559 (2012)
8. Sanchis-Ojeda, R. et al. Transits and occultations of an Earth-sized planet in an 8.5-hour orbit. *Astrophys. J.* **774,** 54–62 (2013)
9. Cosentino, R. et al. Harps-N: the new planet hunter at TNG. *Proc. SPIE* **8446,** 84461V (2012)
10. Mayor, M. et al. Setting new standards with HARPS. *Messenger* **114,** 20–24 (2003)
11. Baranne, A. et al. ELODIE: a spectrograph for accurate radial velocity measurements. *Astron. Astrophys.* **119** (suppl.), 373–390 (1996)
12. Buchhave, L. A. et al. An abundance of small exoplanets around stars with a wide range of metallicities. *Nature* **486,** 375–377 (2012)
13. Yi, S. et al. Toward better age estimates for stellar populations: the $Y^2$ isochrones for solar mixture. *Astrophys. J. Suppl. Ser.* **136,** 417–437 (2001)
14. Lovis, Ch. et al. The HARPS search for southern extra-solar planets. XXXI. Magnetic activity cycles in solar-type stars: statistics and impact on precise radial velocities. Preprint at http://arxiv.org/abs/1107.5325 (2011)
15. Hatzes, A. P. et al. An investigation into the radial velocity variations of CoRoT-7. *Astron. Astrophys.* **520,** A93–A108 (2010)

16. Zechmeister, M. & Kuerster, M. The generalised Lomb-Scargle periodogram. A new formalism for the floating-mean and Keplerian periodograms. *Astron. Astrophys.* **496,** 577–584 (2009)

17. Howard, A. W. *et al.* A rocky composition for an Earth-sized exoplanet. *Nature* http://dx.doi.org/10.1038/nature12767 (this issue).

18. Zeng, L. & Sasselov, D. A detailed model grid for solid planets from 0.1 through 100 Earth masses. *Publ. Astron. Soc. Pacif.* **125,** 227–239 (2013)

19. Rogers, L. A., Bodenheimer, P., Lissauer, J. & Seager, S. The low density limit of the mass-radius relation for exo-Neptunes. *Bull. Am. Astron. Soc.* **43,** 402.04 (2011)

20. Léger, A., Rouan, D. & Schneider, J. Transiting exoplanets from the CoRoT space mission. VIII. CoRoT-7b: the first super-Earth with measured radius. *Astron. Astrophys.* **506,** 287–302 (2009)

21. Marcus, R. A., Sasselov, D., Hernquist, L. & Stewart, S. T. Minimum radii of super-Earths: constraints from giant impacts. *Astrophys. J.* **712,** L73–L76 (2010)

22. Lissauer, J. J. *et al.* All six planets known to orbit Kepler-11 have low densities. *Astrophys. J.* **770,** 131–145 (2013)

23. Charbonneau, D. *et al.* A super-Earth transiting a nearby low-mass star. *Nature* **462,** 891–894 (2009)

24. Smith, J. C. *et al.* Kepler presearch data conditioning II: a Bayesian approach to systematic error correction. *Publ. Astron. Soc. Pacif.* **124,** 1000–1014 (2012)

25. Stumpe, M. C. *et al.* Kepler presearch data conditioning I: architecture and algorithms for error correction in Kepler light curves. *Publ. Astron. Soc. Pacif.* **124,** 985–999 (2012)

26. Schwarz, G. E. Estimating the dimension of a model. *Ann. Stat.* **6,** 461–464 (1978)

27. Liddle, A. R. *Mon. Not. R. Astron. Soc.* **377,** L74–L78 (2007)

28. Burnham, K. P. Multimodel inference: understanding AIC and BIC in model selection. *Sociol. Methods Res.* **33,** 261–304 (2004)

29. Gregory, P. C. *Bayesian Logical Data Analysis for the Physical Sciences* (Cambridge Univ. Press, 2005)

30. Anderson, D. R. *et al.* WASP-30b: a 61 $M_{Jup}$ brown dwarf transiting a $V = 12$, F8 star. *Astrophys J.* **726,** L19-L23 (2011)

# TABLES AND FIGURES

Table 1 - Planetary system parameters of Kepler-78 b as determined from radial velocities BJD_UTC indicates the Barycentric Julian Date expressed in Universal Time UTC. We note that $\lambda_0$ denotes the mean longitude at the mean date of the observing campaign (reference epoch, $T_0$). This coordinate has the advantage of not being degenerate for low eccentricities. Our choice of $T_0$ reduces correlations between adjusted parameters. Stated uncertainties represent the 68.3% confidence interval. (O-C) means the standard deviation of difference between Observed and Computed (modelled) data.

| Planetary system parameter | Unit | Planet b | 68.3% C.I. |
|---|---|---|---|
| Reference epoch $T_0$ | [BJD_UTC] | 2,456,465.076392 | fixed |
| *Orbital inclination $i$ | [deg] | 79 | -14 +9 |
| †Systemic radial velocity $\gamma$ | [km s$^{-1}$] | -3.5084 | ±0.0008 |
| †Orbital period $P$ | [days] | 0.3550 | ±0.0004 |
| †Mean longitude $\lambda_0$ at epoch $T_0$ | [deg] | 293 | ±13 |
| †Eccentricity $e$ |  | 0 | fixed |
| †Radial-velocity semi-amplitude $K_p$ | [m s$^{-1}$] | 1.96 | ±0.32 |
| ‡Planetary mass $m_p$ | [M$_\oplus$] | 1.86 | -0.25 +0.38 |
| ‡Planetary radius $R_p$ | [R$_\oplus$] | 1.173 | -0.089 +0.159 |
| ‡Planetary mean density $\rho_p$ | [g cm$^{-3}$] | 5.57 | -1.31 +3.02 |
| ‡Semi-major axis $a$ | [AU] | 0.0089 | - |
| *Surface temperature $T$ | [K] | 1500-3000 | - |
| Number of measurements $N_{meas}$ |  | 109 | - |
| Time span of observations | [days] | 97.1 | - |
| †Radial-velocity dispersion (O-C) | [m s$^{-1}$] | 2.34 | - |
| †Reduced chi-square $\chi^2$ |  | 1.12 | ±0.07 |

*Taken from the discovery paper[8]. †Fitted orbital parameters (maximum likelihood). ‡Mode and confidence interval of the real distribution deduced from the MCMC analysis.

Figure 1 - Radial velocities of Kepler-78 as a function of time. The error bars indicate the estimated internal error (mainly photon-noise-induced error), which was ~2.3 m s$^{-1}$ on average. The signal is dominated by stellar effects. The raw radial-velocity dispersion is 4.08 m s$^{-1}$, which is about twice the photon noise. JD, Julian date.

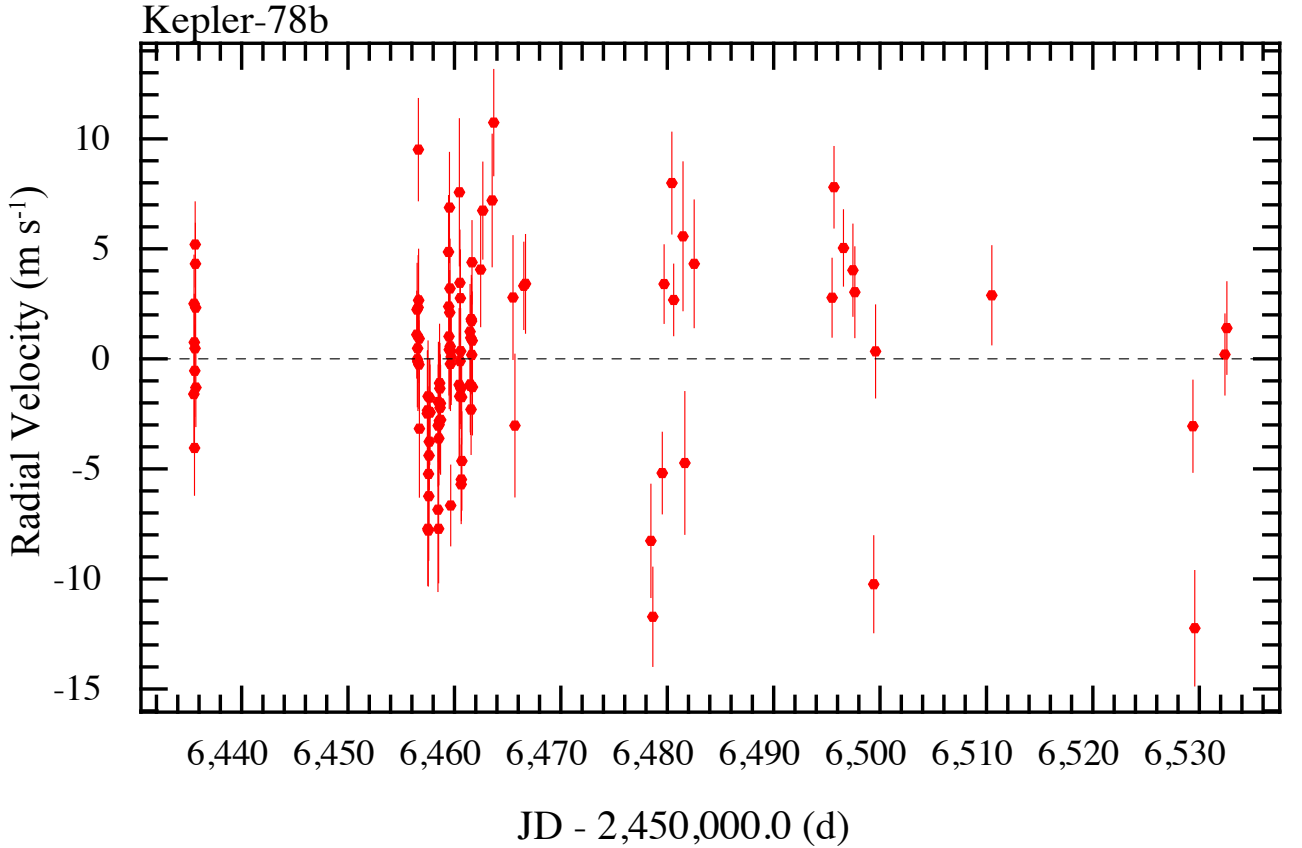

Figure 2 - Model fit of the radial velocities of Kepler-78. a, Grey-scale contour plot of the $\chi^2$ surface around the ephemeris for the period and mid-time of transit from ref. (origin of plot). $\Delta P$ and $\Delta t_0$ represent the departure from the expected ephemeris in units of days. The position of the minimum of the residuals perfectly matches the expected values. b, Phase-folded radial velocities and fitted Keplerian orbit of the signal induced by Kepler-78b after removal of the modelled stellar noise components. $\lambda$, mean longitude. The transit occurs at $\lambda_T = 90°$. We note that the higher number of data points at maximum and minimum radial velocity is a direct consequence of our strategy of observing at quadrature, where the information on amplitude is highest. The red dots show the radial velocities and the corresponding errors when binned over the orbital phase. The error bars of the individual measurements indicate the estimated internal errors (photon-noise-induced error), of ~2.3 m s$^{-1}$ on average. The error bars on the averaged data points essentially represent the internal error of an individual measurement divided by the square root of the number of averaged measurements.

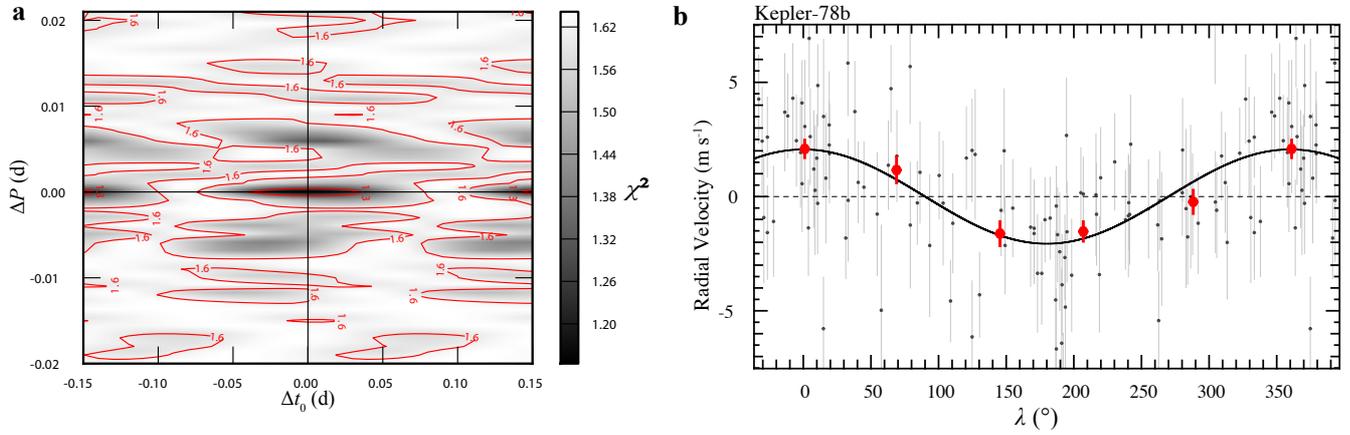

Figure 3 - The hot, rocky planet Kepler-78b placed on a planetary mass–radius diagram. Here masses and radii represent the modes of the corresponding distributions derived from the MCMC analysis, and the error bars represent the 68.3% confidence interval (1$\sigma$). For comparison, Earth and Venus are indicated in the same diagram by corresponding symbols. The other exoplanets shown are those for which mass and radius have been estimated (K, Kepler; Co, CoRoT). From top to bottom, the solid lines show mass and radius for planets consisting of pure water, 50% water, 100% silicates, 50% silicates and 50% iron core, and 100% iron, as computed with the theoretical models of ref. 18. The dashed line shows the maximum mantle-stripping models from ref. 21; that is, Earth-like exoplanets made of pure iron are not expected to form around normal stars. From top to bottom, the dotted lines represent mean densities of 1, 2, 4 and 8 g cm$^{-3}$.

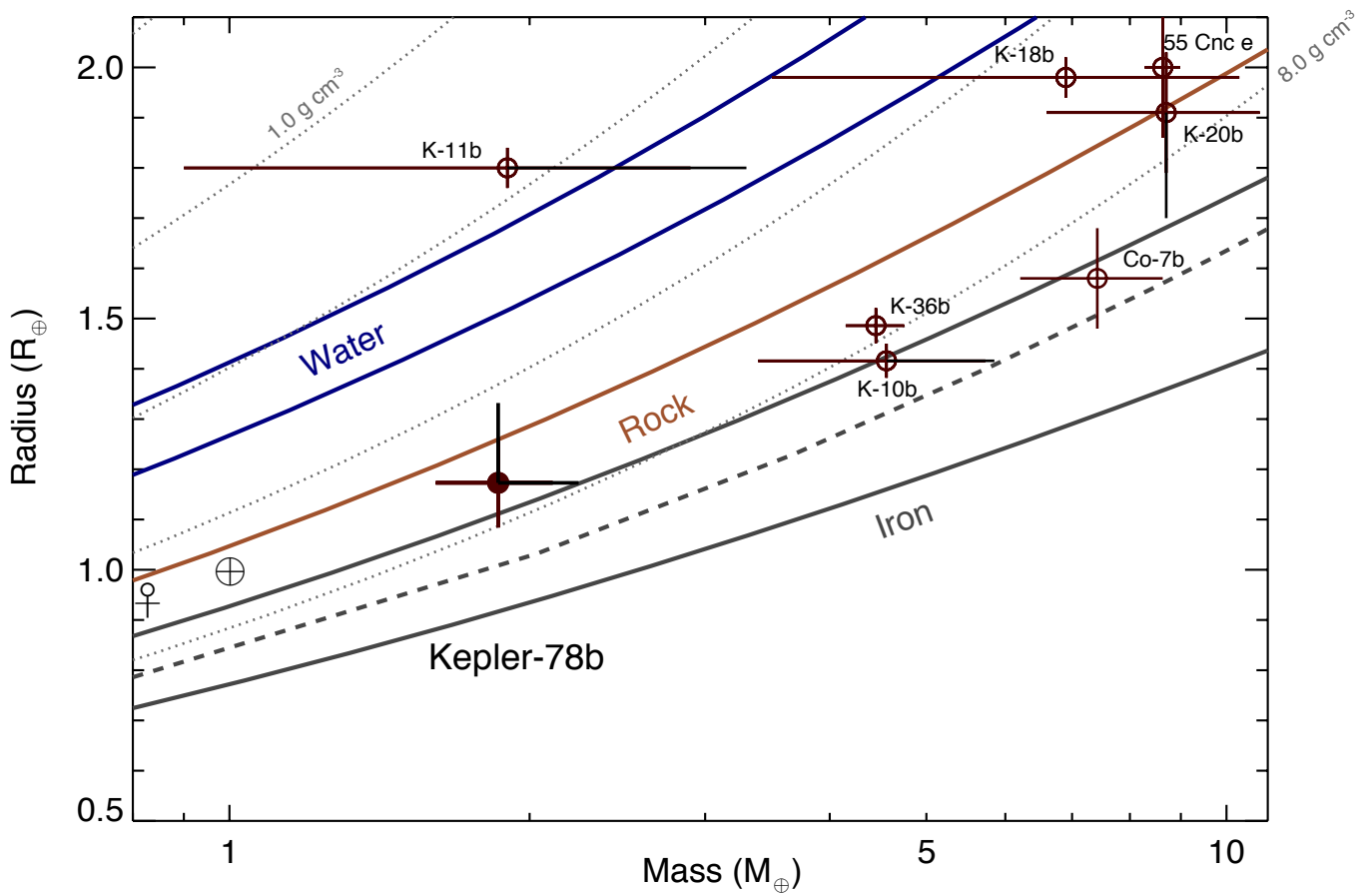

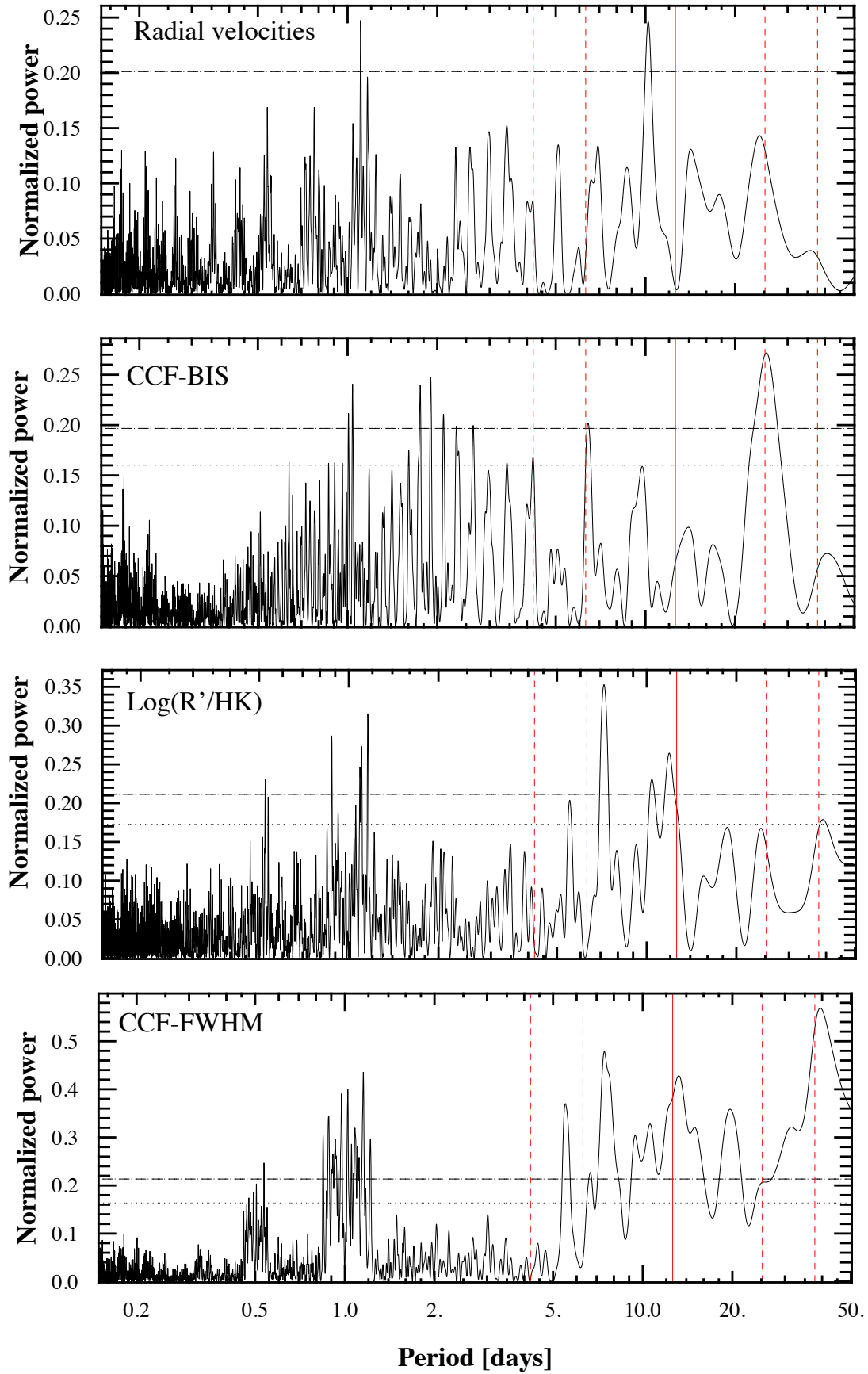

**Extended Data Figure 1** - Generalized Lomb–Scargle periodogram of several parameters measured by HARPS-N. The panels show, from top to bottom, the periodogram of the radial velocities (RV) of Kepler-78, the line bisector (CCF-BIS), the activity indicator ($\log(R'_{HK})$) and the full-width at half maximum (CCF-FWHM) of Kepler-78. The dotted and dashed horizontal lines represent the 10% and 1% false-alarm probabilities, respectively. The vertical lines show the stellar rotational period (solid) and its two first harmonics (dashed). All the indicators show excess energy at periods of around 6 d and above, indicating that the peak observed in the radial-velocity data at a period of about 10 d is most likely to have a stellar origin. The additional power in the line bisector periodogram at periods longer than 1 d is most probably induced by stellar spots.

**Extended Data Figure 2** - Kepler light curve of Kepler-78d. The data have been de-trended using the PDC-MAP algorithm. Different colours represent different quarters of observation.

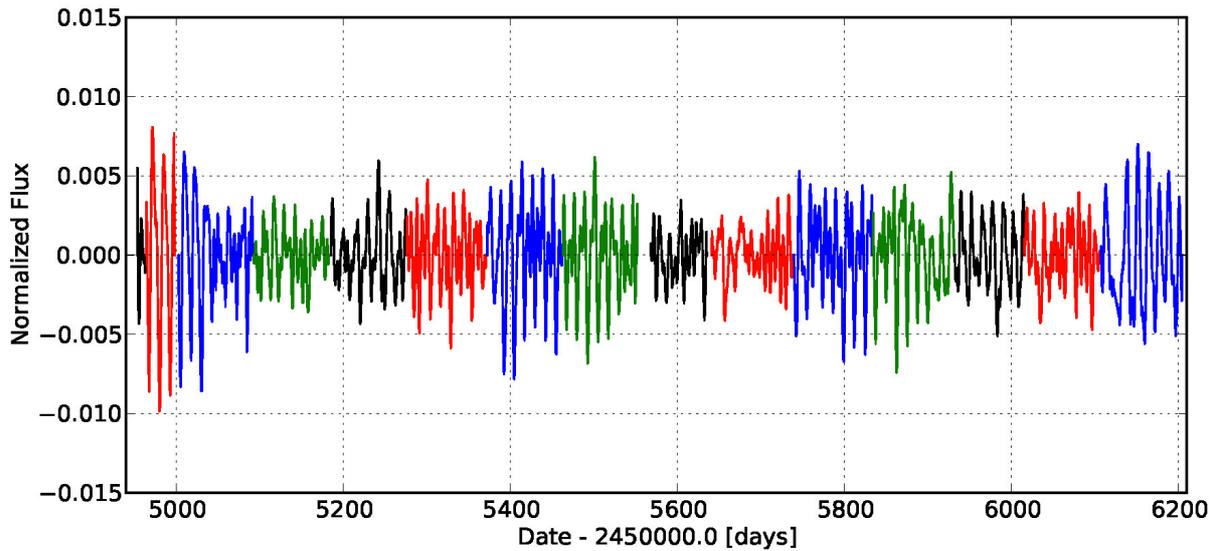

**Extended Data Figure 3** - Spectral analysis of the Kepler light curve. a, ACF of the Kepler light curve showing correlation peaks every 12.6 d and a decay on an e-folding timescale of ~50 d. b, Power spectral distribution of the Kepler light curve. Peaks are well identified at the stellar rotational period of 12.6 d and its two first harmonics. At shorter periods, the signal and several harmonics of the transiting planet Kepler-78b can be identified.

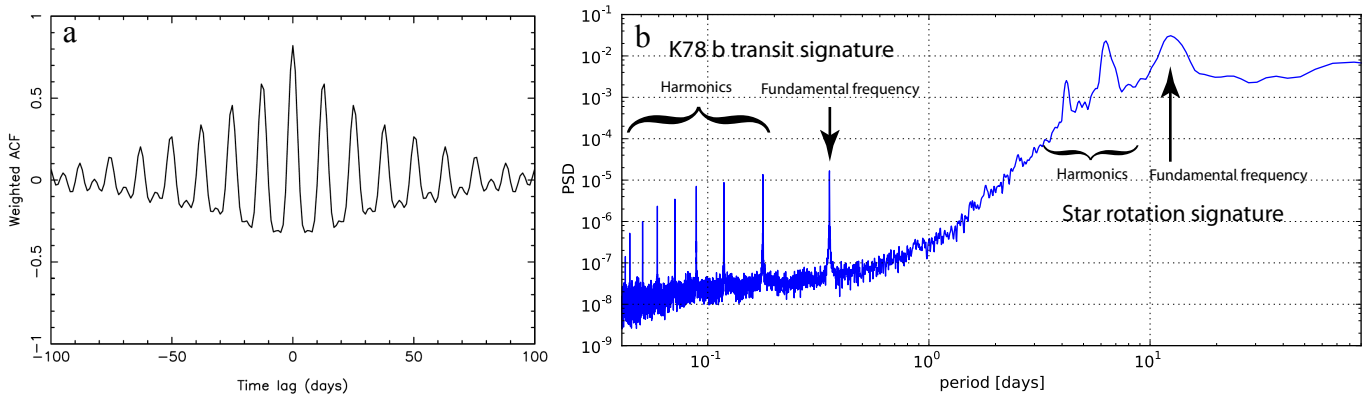

**Extended Data Figure 4** - Periodogram of the radial-velocity residuals after subtraction of the 4.2-d and 10.0-d stellar components. The dotted and dashed horizontal lines represent the 10% and 1% false-alarm probabilities, respectively. The signature of Kepler-78b (and its aliases) can now clearly be identified with a false-alarm probability significantly lower than 1%.

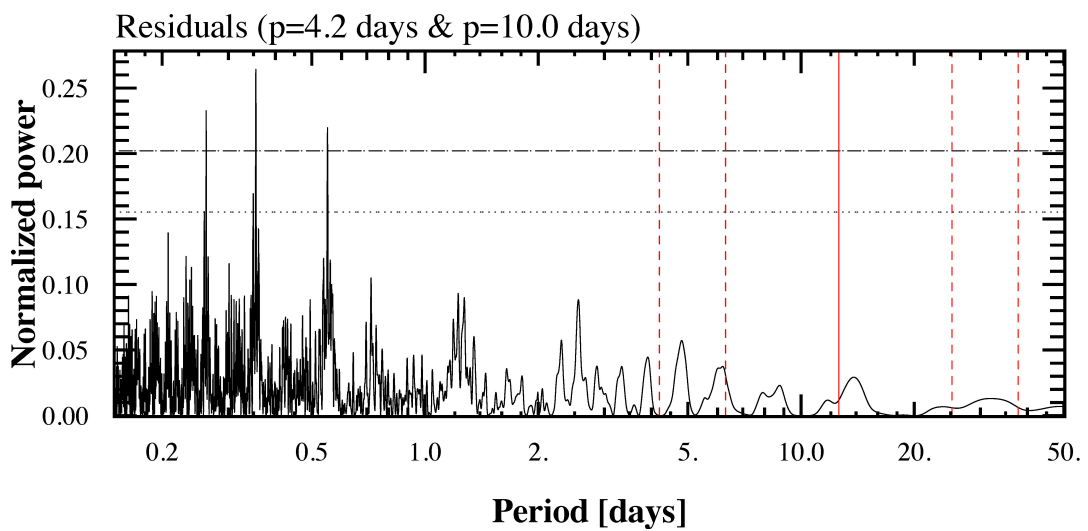

**Extended Data Figure 5 -** Probability density functions derived from the MCMC analysis. a, Probability density function of the planetary density. b, Probability density function of the planetary mass.

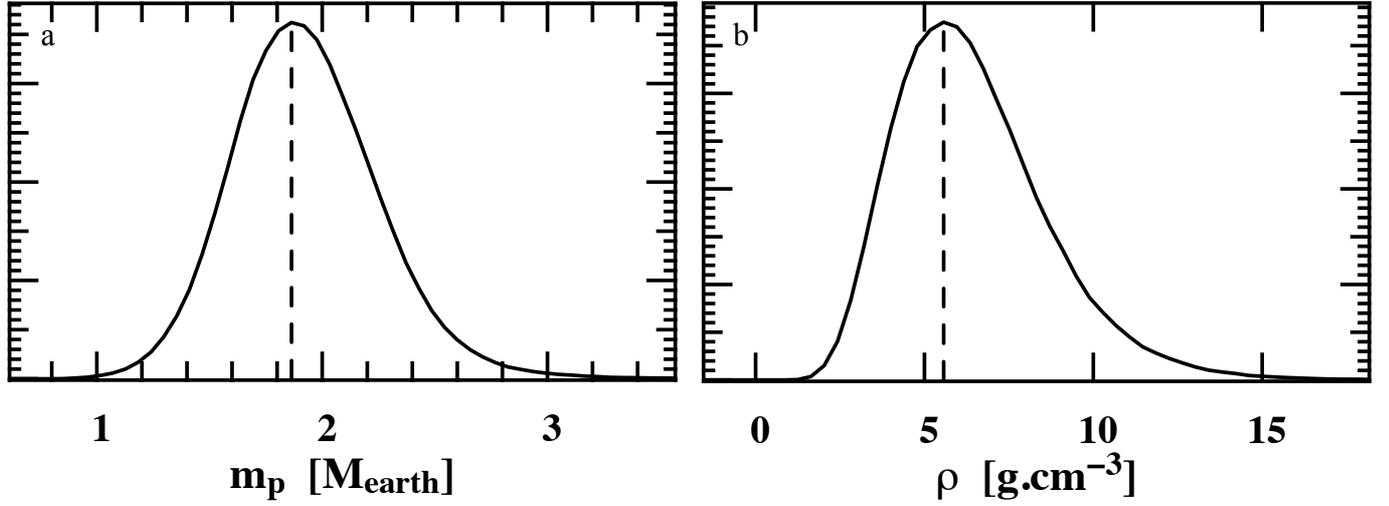

**Extended Data Table 1 -** Stellar parameters of Kepler-78 computed from HARPS-N spectra

| Stellar parameter | Unit | Value | 68.3% C.I. |
|---|---|---|---|
| *Effective temperature, $T_{eff}$ | [K] | 5058 | ±50 |
| *Surface gravity, log g | [g in cm s$^{-2}$] | 4.55 | ±0.10 |
| *Metallicity, [m/H] | | -0.18 | ±0.08 |
| †Mass of the star, $M_*$ | [M$_\odot$] | 0.758 | ±0.046 |
| †Radius of the star, $R_*$ | [R$_\odot$] | 0.737 | +0.034 -0.042 |
| ‡Ca II H&K activity indicator, log($R'_{HK}$) | | -4.52 | ±0.015§ |
| ‡Projected rotational velocity, $v \sin i$ | [km s$^{-1}$] | 2.8 | ±0.5 |

*Obtained from an SPC analysis of the HARPS-N spectra. †Based on the Yonsei–Yale stellar evolutionary models[13]. ‡Output of the HARPS-N data-reduction pipeline using $B-V$ = 0.91. §The error indicates the standard deviation of the individual values.

**Extended Data Table 2 -** Comparison of the statistical 'quality' of all the considered models Model 5 with the three Keplerians is clearly the one best representing the data, given the fact that its BIC evidence ratio, ER, is lowest and its BIC weight, w, highest. For comparison, the $\chi^2$ and the reduced $\chi^2$ of the residuals are also given. It is interesting to note that the most likely model does not necessarily have the lowest reduced $\chi^2$. $N_{par}$ indicates the number of free parameters in the model.

| Model | Description | Stellar Jitter | Planet | X2 | $\chi^2$ | $N_{par}$ | ER | ΔER | w |
|---|---|---|---|---|---|---|---|---|---|
| 1 | D0 | No | None | 374.38 | 3.47 | 1 | 379.07 | 209.2 | 0.00 |
| 2 | D0 + K1 | No | K78 (fixed) | 326.43 | 3.05 | 2 | 335.81 | 166.0 | 0.00 |
| 3 | D0 + nightly offsets | 21 offsets | K78 (fixed) | 89.90 | 1.03 | 22 | 193.11 | 23.3 | 0.00 |
| 4 | D0 + K2 | 2 Keplerians (P2=4.2; e2=0; P3=10.2; e3 adj) | None | 166.03 | 1.69 | 11 | 217.63 | 47.8 | 0.00 |
| 5 | D0 + K3 | 2 Keplerians (P2=4.2; e2=0; P3=10.2; e3 adj) | K78 (fixed) | 122.94 | 1.24 | 10 | **169.85** | 0.00 | **0.710** |
| 6 | D0 + K4 | 3 Keplerians (P2=4.2; e2=0; P3=6.5; e3 adj; P4=23.58; e4=0) | K78 (fixed) | 120.40 | 1.23 | 11 | 172.00 | 2.2 | 0.242 |

D0 represents a constant term, K$n$ is the number of Keplerians in the model. $ei$ and $Pi$ respectively indicate the eccentricity and the period of the $i$th planet.

Extended Data Table 3 - Orbital parameters (distributions) of the planet and parameters of the two additional Keplerians describing the star-induced signal as determined from the MCMC analysis. $P_i$ represents the (orbital) period, $K$ the semi-amplitude, $e_i$ the eccentricity and $\lambda_{0i}$ the mean longitude of the $i$th signal at the reference epoch, $T_0$. The index $i$ designates the planet (p) and the two additional (stellar) signals (2 and 3). $m_p\sin(i)$ is the minimum mass of the planet and $a_p$ is the semi-major axis of the orbit.

| Parameters | Unit | Distribution mode | 99% Confidence Interval |
|---|---|---|---|
| $P_p$ | [days] | 0.35500743 | 0.35500729 - 0.35500760 |
| $K_p$ | [m s$^{-1}$] | 1.986 | 1.243 - 2.679 |
| $e_p$ | | 0 | fixed |
| $\lambda_{0p}$ | [deg] | 297.53 | 296.82 - 298.23 |
| $T_{0p}$ | [days] | 1465.165144271 | 1465.165144235 - 1465.165144313 |
| $m_p \sin(i)$ | [M$_\oplus$] | 1.82 | 1.12 - 2.51 |
| $a_p$ | [10$^{-3}$ AU] | 8.95323 | 8.448 - 9.393 |
| $P_2$ | [days] | 4.210 | 4.193 - 4.237 |
| $K_2$ | [m s-1] | 3.98 | 2.66 - 5.29 |
| $e_2$ | | 0 | fixed |
| $\lambda_{02}$ | [deg] | 146 | 118 -166 |
| $P_3$ | [days] | 10.037 | 9.977 - 10.272 |
| $K_3$ | [m s-1] | 5.33 | 4.14 - 6.78 |
| $e_3$ | | 0.39 | 0.03 - 0.65 |
| $\lambda_{03}$ | [deg] | -320.0 | -329 - -310 |
| $\omega_3$ | [deg] | 154 | 91 - 262 |

Extended Data Table 4 - Planetary parameters derived from the MCMC analysis. The table gives the distributions of the mass, radius and density of Kepler-78b.

| Statistical parameter | Mass [M$_\oplus$] | Radius [R$_\oplus$] | Density [g cm$^{-3}$] |
|---|---|---|---|
| Mode | 1.86 | 1.173 | 5.57 |
| Median | 1.91 | 1.194 | 6.13 |
| 68.3% confidence interval | 1.61 – 2.24 | 1.084 – 1.332 | 4.26 – 8.59 |
| 99% confidence interval | 1.17 – 3.00 | 0.942 – 1.584 | 2.34 – 14.29 |

## SUPPLEMENTARY DATA

HARPS-N Data for Kepler-78b. From left to right are given: Julian Date, Radial Velocity $RV$, the estimated error $\sigma_{RV}$ on the radial velocity, the Full-Width Half Maximum ($FWHM$) and the line-bisector ($BIS$) of the Cross-Correlation Function (CCF), the CaII activity indicator log($R'_{HK}$) and its error $\sigma_{\log(R'HK)}$.

| Julian Date [Day] BJD_UTC | $RV$ [km s$^{-1}$] | $\sigma_{RV}$ [km s$^{-1}$] | $FWHM$ [km s$^{-1}$] | $BIS$ [km s$^{-1}$] | log($R'_{HK}$) | $\sigma_{\log(R'HK)}$ |
|---|---|---|---|---|---|---|
| 2'456'435.51292 | -3.51192 | 0.00246 | 6.95372 | 0.01851 | -4.5398 | 0.0115 |
| 2'456'435.53406 | -3.50781 | 0.00220 | 6.94868 | 0.01863 | -4.5175 | 0.0090 |
| 2'456'435.55521 | -3.50957 | 0.00218 | 6.95888 | 0.01731 | -4.5233 | 0.0088 |
| 2'456'435.57634 | -3.51437 | 0.00214 | 6.95906 | 0.01969 | -4.5269 | 0.0084 |
| 2'456'435.59748 | -3.51086 | 0.00207 | 6.96806 | 0.02241 | -4.5146 | 0.0077 |
| 2'456'435.61863 | -3.50984 | 0.00207 | 6.95702 | 0.02996 | -4.5087 | 0.0076 |
| 2'456'435.63976 | -3.50512 | 0.00193 | 6.96108 | 0.02084 | -4.5030 | 0.0067 |
| 2'456'435.66092 | -3.50600 | 0.00184 | 6.95719 | 0.02350 | -4.5085 | 0.0063 |
| 2'456'435.68207 | -3.50799 | 0.00190 | 6.95936 | 0.02604 | -4.5108 | 0.0067 |
| 2'456'435.70320 | -3.51162 | 0.00177 | 6.96916 | 0.02262 | -4.5221 | 0.0062 |
| 2'456'456.46507 | -3.50922 | 0.00197 | 6.95166 | 0.01866 | -4.5299 | 0.0074 |
| 2'456'456.48977 | -3.50808 | 0.00210 | 6.95563 | 0.02374 | -4.5267 | 0.0080 |
| 2'456'456.51440 | -3.51034 | 0.00215 | 6.96748 | 0.02421 | -4.5031 | 0.0077 |
| 2'456'456.53902 | -3.50984 | 0.00186 | 6.95824 | 0.02026 | -4.5008 | 0.0061 |
| 2'456'456.56315 | -3.51045 | 0.00221 | 6.95990 | 0.02738 | -4.5153 | 0.0082 |
| 2'456'456.58768 | -3.50798 | 0.00235 | 6.94980 | 0.02407 | -4.5115 | 0.0087 |
| 2'456'456.61247 | -3.50081 | 0.00232 | 6.96344 | 0.01740 | -4.5245 | 0.0088 |
| 2'456'456.63656 | -3.50766 | 0.00232 | 6.95182 | 0.01921 | -4.5169 | 0.0087 |

| Julian Date [Day] BJD_UTC | RV [km s$^{-1}$] | $\sigma_{RV}$ [km s$^{-1}$] | FWHM [km s$^{-1}$] | BIS [km s$^{-1}$] | log($R'_{HK}$) | $\sigma_{\log(R'HK)}$ |
|---|---|---|---|---|---|---|
| 2'456'456.66187 | -3.51058 | 0.00199 | 6.96419 | 0.01943 | -4.5207 | 0.0070 |
| 2'456'456.68591 | -3.50940 | 0.00199 | 6.95225 | 0.02018 | -4.5240 | 0.0069 |
| 2'456'456.70988 | -3.51349 | 0.00311 | 6.95434 | 0.02068 | -4.5016 | 0.0128 |
| 2'456'457.45301 | -3.51280 | 0.00283 | 6.94809 | 0.03428 | -4.5101 | 0.0124 |
| 2'456'457.47415 | -3.51266 | 0.00271 | 6.94241 | 0.03203 | -4.5229 | 0.0118 |
| 2'456'457.49848 | -3.51804 | 0.00257 | 6.94528 | 0.03760 | -4.5429 | 0.0112 |
| 2'456'457.51963 | -3.51202 | 0.00251 | 6.96096 | 0.03207 | -4.5414 | 0.0107 |
| 2'456'457.5436 | -3.51812 | 0.00252 | 6.95148 | 0.03068 | -4.5228 | 0.0103 |
| 2'456'457.45673 | -3.51555 | 0.00267 | 6.95343 | 0.03531 | -4.5333 | 0.0113 |
| 2'456'457.58899 | -3.51656 | 0.00291 | 6.96189 | 0.03625 | -4.5357 | 0.0129 |
| 2'456'457.61014 | -3.51471 | 0.00312 | 6.96346 | 0.03787 | -4.5468 | 0.0146 |
| 2'456'457.63537 | -3.51408 | 0.00218 | 6.94145 | 0.03292 | -4.5149 | 0.0081 |
| 2'456'457.65652 | -3.51208 | 0.00175 | 6.95862 | 0.03767 | -4.5118 | 0.0057 |
| 2'456'457.68068 | -3.51273 | 0.00239 | 6.95479 | 0.03651 | -4.5194 | 0.0094 |
| 2'456'457.70181 | -3.51276 | 0.00216 | 6.94697 | 0.03708 | -4.5113 | 0.0079 |
| 2'456'458.44957 | -3.51717 | 0.00372 | 6.94501 | 0.03106 | -4.5053 | 0.0184 |
| 2'456'458.47071 | -3.51335 | 0.00275 | 6.93871 | 0.02770 | -4.5165 | 0.0121 |
| 2'456'458.49468 | -3.51228 | 0.00271 | 6.94782 | 0.02794 | -4.5092 | 0.0113 |
| 2'456'458.51583 | -3.51804 | 0.00245 | 6.95567 | 0.03053 | -4.5288 | 0.0100 |
| 2'456'458.53986 | -3.51393 | 0.00211 | 6.94653 | 0.02689 | -4.5338 | 0.0080 |
| 2'456'458.56099 | -3.51331 | 0.00219 | 6.93902 | 0.02737 | -4.5142 | 0.0081 |
| 2'456'458.58499 | -3.51313 | 0.00219 | 6.94344 | 0.02798 | -4.5237 | 0.0082 |
| 2'456'458.60612 | -3.51142 | 0.00267 | 6.95382 | 0.03788 | -4.5055 | 0.0107 |
| 2'456'458.63003 | -3.51166 | 0.00204 | 6.93839 | 0.02192 | -4.5264 | 0.0074 |
| 2'456'458.65118 | -3.51254 | 0.00198 | 6.94624 | 0.03881 | -4.5164 | 0.0070 |
| 2'456'458.68064 | -3.51234 | 0.00220 | 6.94080 | 0.02335 | -4.5137 | 0.0081 |
| 2'456'458.70178 | -3.51309 | 0.00247 | 6.93344 | 0.02536 | -4.5297 | 0.0100 |
| 2'456'459.44426 | -3.50546 | 0.00256 | 6.93468 | 0.02975 | -4.5231 | 0.0112 |
| 2'456'459.4654 | -3.50794 | 0.00255 | 6.94565 | 0.03123 | -4.5314 | 0.0110 |
| 2'456'459.49047 | -3.50930 | 0.00266 | 6.93171 | 0.02921 | -4.5244 | 0.0114 |
| 2'456'459.51161 | -3.50990 | 0.00268 | 6.93251 | 0.01346 | -4.5289 | 0.0113 |
| 2'456'459.53456 | -3.50344 | 0.00250 | 6.94324 | 0.02906 | -4.5219 | 0.0100 |
| 2'456'459.55680 | -3.50821 | 0.00190 | 6.93487 | 0.02634 | -4.5246 | 0.0066 |
| 2'456'459.58083 | -3.50712 | 0.00224 | 6.94046 | 0.02936 | -4.5306 | 0.0085 |
| 2'456'459.60197 | -3.50975 | 0.00198 | 6.93618 | 0.02662 | -4.5249 | 0.0072 |
| 2'456'459.62610 | -3.51055 | 0.00211 | 6.93960 | 0.03224 | -4.5406 | 0.0081 |
| 2'456'459.64723 | -3.51698 | 0.00183 | 6.93684 | 0.01974 | -4.5257 | 0.0063 |
| 2'456'459.67153 | -3.50997 | 0.00222 | 6.93272 | 0.02136 | -4.5223 | 0.0084 |
| 2'456'459.69267 | -3.51023 | 0.00216 | 6.93756 | 0.02153 | -4.5238 | 0.0080 |
| 2'456'460.44685 | -3.51150 | 0.00346 | 6.91362 | 0.02452 | -4.5485 | 0.0173 |
| 2'456'460.46800 | -3.50275 | 0.00334 | 6.92454 | 0.03807 | -4.5403 | 0.0161 |
| 2'456'460.49402 | -3.51202 | 0.00285 | 6.93892 | 0.02922 | -4.5365 | 0.0127 |
| 2'456'460.51515 | -3.50686 | 0.00239 | 6.93383 | 0.02125 | -4.5199 | 0.0093 |
| 2'456'460.53921 | -3.51042 | 0.00283 | 6.93975 | 0.03841 | -4.5555 | 0.0129 |
| 2'456'460.56036 | -3.50756 | 0.00219 | 6.94325 | 0.02336 | -4.5244 | 0.0082 |
| 2'456'460.58477 | -3.50997 | 0.00193 | 6.93332 | 0.02479 | -4.5346 | 0.0068 |
| 2'456'460.60592 | -3.51167 | 0.00181 | 6.92977 | 0.03427 | -4.5268 | 0.0061 |

| Julian Date [Day] BJD_UTC | RV [km s$^{-1}$] | $\sigma_{RV}$ [km s$^{-1}$] | FWHM [km s$^{-1}$] | BIS [km s$^{-1}$] | log($R'_{HK}$) | $\sigma_{\log(R'HK)}$ |
|---|---|---|---|---|---|---|
| 2'456'460.63011 | -3.51602 | 0.00178 | 6.92925 | 0.02366 | -4.5299 | 0.0061 |
| 2'456'460.65127 | -3.51580 | 0.00186 | 6.93402 | 0.03138 | -4.5221 | 0.0064 |
| 2'456'460.67548 | -3.51206 | 0.00212 | 6.94082 | 0.03032 | -4.5305 | 0.0079 |
| 2'456'460.69663 | -3.51496 | 0.00223 | 6.93131 | 0.01716 | -4.5234 | 0.0083 |
| 2'456'461.47151 | -3.50908 | 0.00214 | 6.92998 | 0.02889 | -4.5455 | 0.0085 |
| 2'456'461.49265 | -3.51148 | 0.00215 | 6.93009 | 0.02596 | -4.5418 | 0.0084 |
| 2'456'461.51660 | -3.51155 | 0.00220 | 6.93285 | 0.02729 | -4.5403 | 0.0086 |
| 2'456'461.53775 | -3.50936 | 0.00192 | 6.93311 | 0.03235 | -4.5385 | 0.0069 |
| 2'456'461.56161 | -3.51262 | 0.00204 | 6.93480 | 0.03239 | -4.5481 | 0.0078 |
| 2'456'461.58275 | -3.50851 | 0.00198 | 6.93349 | 0.02517 | -4.5391 | 0.0073 |
| 2'456'461.60669 | -3.50860 | 0.00193 | 6.93417 | 0.02476 | -4.5360 | 0.0071 |
| 2'456'461.62782 | -3.51014 | 0.00176 | 6.93482 | 0.02080 | -4.5423 | 0.0062 |
| 2'456'461.65172 | -3.50593 | 0.00189 | 6.93816 | 0.02195 | -4.5288 | 0.0067 |
| 2'456'461.67287 | -3.50949 | 0.00221 | 6.93511 | 0.01626 | -4.5240 | 0.0085 |
| 2'456'461.69683 | -3.51160 | 0.00217 | 6.92724 | 0.03118 | -4.5444 | 0.0085 |
| 2'456'462.46355 | -3.50626 | 0.00259 | 6.93986 | 0.02296 | -4.5024 | 0.0106 |
| 2'456'462.66329 | -3.50358 | 0.00220 | 6.93904 | 0.01201 | -4.5071 | 0.0081 |
| 2'456'463.54683 | -3.50312 | 0.00301 | 6.95318 | 0.00594 | -4.5173 | 0.0135 |
| 2'456'463.69612 | -3.49958 | 0.00241 | 6.95904 | 0.01523 | -4.5138 | 0.0096 |
| 2'456'465.50148 | -3.50753 | 0.00281 | 6.95763 | 0.03446 | -4.5123 | 0.0116 |
| 2'456'465.68396 | -3.51335 | 0.00324 | 6.97210 | 0.04367 | -4.5200 | 0.0144 |
| 2'456'466.52111 | -3.50700 | 0.00198 | 6.93773 | 0.03711 | -4.5134 | 0.0069 |
| 2'456'466.67679 | -3.50691 | 0.00224 | 6.94697 | 0.02955 | -4.5311 | 0.0088 |
| 2'456'478.46010 | -3.51859 | 0.00257 | 6.99142 | 0.05363 | -4.5142 | 0.0104 |
| 2'456'478.64282 | -3.52204 | 0.00225 | 6.98054 | 0.05491 | -4.4990 | 0.0082 |
| 2'456'479.52887 | -3.51551 | 0.00185 | 6.96010 | 0.03046 | -4.5118 | 0.0062 |
| 2'456'479.70375 | -3.50692 | 0.00178 | 6.96035 | 0.03277 | -4.5023 | 0.0057 |
| 2'456'480.43572 | -3.50233 | 0.00231 | 6.97371 | 0.02707 | -4.5330 | 0.0094 |
| 2'456'480.60732 | -3.50764 | 0.00162 | 6.96802 | 0.02679 | -4.5201 | 0.0052 |
| 2'456'481.48297 | -3.50475 | 0.00338 | 6.97512 | 0.02492 | -4.5420 | 0.0166 |
| 2'456'481.65660 | -3.51505 | 0.00324 | 6.96638 | 0.03568 | -4.5485 | 0.0158 |
| 2'456'482.53436 | -3.50600 | 0.00290 | 6.94084 | 0.03284 | -4.5526 | 0.0136 |
| 2'456'495.49348 | -3.50754 | 0.00179 | 6.95661 | 0.04431 | -4.5443 | 0.0061 |
| 2'456'495.66755 | -3.50252 | 0.00185 | 6.96448 | 0.03842 | -4.5329 | 0.0063 |
| 2'456'496.56297 | -3.50528 | 0.00173 | 6.95914 | 0.03905 | -4.5356 | 0.0058 |
| 2'456'497.44726 | -3.50629 | 0.00209 | 6.94333 | 0.04797 | -4.5339 | 0.0079 |
| 2'456'497.62894 | -3.50729 | 0.00206 | 6.93657 | 0.03749 | -4.5322 | 0.0078 |
| 2'456'499.40721 | -3.52056 | 0.00220 | 6.91500 | 0.02867 | -4.5449 | 0.0089 |
| 2'456'499.57787 | -3.50998 | 0.00211 | 6.91118 | 0.03103 | -4.5420 | 0.0080 |
| 2'456'510.50035 | -3.50743 | 0.00225 | 6.95198 | 0.03333 | -4.5628 | 0.0098 |
| 2'456'529.39927 | -3.51338 | 0.00209 | 6.98691 | 0.04629 | -4.4790 | 0.0070 |
| 2'456'529.57873 | -3.52256 | 0.00262 | 6.97892 | 0.03428 | -4.4799 | 0.0100 |
| 2'456'532.41047 | -3.51012 | 0.00184 | 6.95228 | 0.02116 | -4.5322 | 0.0065 |
| 2'456'532.58840 | -3.50892 | 0.00210 | 6.96228 | 0.03283 | -4.5221 | 0.0081 |